\documentclass[reprint,
superscriptaddress,
amsmath,amssymb,prb]{revtex4-2}
\usepackage{graphicx}
\usepackage{xcolor}
\usepackage{soul}
\usepackage{newtxtext}
\usepackage[cmintegrals]{newtxmath} 
\usepackage{graphicx}
\usepackage{dcolumn}
\usepackage{bm}
\usepackage{hyperref}
\usepackage{amsmath}
\usepackage{cleveref}
\hypersetup{colorlinks,
	linkcolor={blue!75!black!80!yellow},
	citecolor={blue!75!black!80!yellow},
	urlcolor={blue!75!black!80!yellow}
}

\makeatletter \renewcommand\@make@capt@title[2]{%
\@ifx@empty\float@link{\@firstofone}{\expandafter\href\expandafter{\float@link}}%
\sffamily{\textbf{#1}}\@caption@fignum@sep#2 }

\usepackage[normalem]{ulem}
\makeatletter
\renewcommand\@make@capt@title[2]{%
    \@ifx@empty\float@link{\@firstofone}{\expandafter\href\expandafter{\float@link}}%
    \sffamily{\textbf{#1}}\@caption@fignum@sep#2
}%

\begin{document}

\preprint{APS/123-QED}

\title{Finite-size effects of electron transport in PdCoO$_2$}

\author{Georgios Varnavides}
\thanks{These authors contributed equally.}
\affiliation{John A. Paulson School of Engineering and Applied Sciences, Harvard University, Cambridge, MA 02138, USA}
\affiliation{Department of Materials Science and Engineering, Massachusetts Institute of Technology, Cambridge, MA 02139, USA}
\affiliation{Research Laboratory of Electronics, Massachusetts Institute of Technology, Cambridge, MA 02139, USA}
\author{Yaxian Wang}
\thanks{These authors contributed equally.}
\affiliation{John A. Paulson School of Engineering and Applied Sciences, Harvard University, Cambridge, MA 02138, USA}
\author{Philip J.W. Moll}
\affiliation{Laboratory of Quantum Materials (QMAT), Institute of Materials (IMX), École Polytechnique Fédérale de Lausanne (EPFL), 1015 Lausanne, Switzerland}
\author{Polina Anikeeva}
\affiliation{Department of Materials Science and Engineering, Massachusetts Institute of Technology, Cambridge, MA 02139, USA}
\affiliation{Research Laboratory of Electronics, Massachusetts Institute of Technology, Cambridge, MA 02139, USA}
\author{Prineha Narang}
\email[Electronic address:\;]{prineha@seas.harvard.edu}
\affiliation{John A. Paulson School of Engineering and Applied Sciences, Harvard University, Cambridge, MA 02138, USA}

\date{\today}

\begin{abstract}
A wide range of unconventional transport phenomena have recently been observed in single-crystal delafossite metals.
Here, we present a theoretical framework to elucidate electron transport using a combination of first-principles calculations and numerical modeling of the anisotropic Boltzmann transport equation.
Using PdCoO$_2$ as a model system, we study different microscopic electron and phonon scattering mechanisms and establish the mean free path hierarchy of quasiparticles at different temperatures.
We treat the anisotropic Fermi surface explicitly to numerically obtain experimentally-accessible transport observables, which bridge between the ``diffusive'', ``ballistic'', and ``hydrodynamic'' transport regime limits.
We illustrate that distinction between the ``quasi-ballistic'', and ``quasi-hydrodynamic'' regimes is challenging and often needs to be quantitative in nature.
From first-principles calculations, we populate the resulting transport regime plots, and demonstrate how the Fermi surface orientation adds complexity to the observed transport signatures in micro-scale devices.
Our work provides key insights into microscopic interaction mechanisms on open hexagonal Fermi surfaces and establishes their connection to the macroscopic electron transport in finite-size channels. 
\end{abstract}

\maketitle
Signatures of unconventional transport in condensed matter systems, have been the subject of active research in fields ranging from materials physics to computational hydrodynamics and statistical physics.
Despite the success of Mott's relation in describing the dynamics of charge transport in bulk metals and semimetals at and near the Fermi level~\cite{cutler1969observation}, finite-size effects and the interplay between the geometrical and topological character of the Fermi surface can give rise to unconventional transport regimes.
Of particular interest, is the electron `hydrodynamic' transport regime, which manifests itself in negative non-local resistance and non-uniform current distributions~\cite{gurzhi1968hydrodynamic,jong1995hydrodynamic,levitov_electron_2016}, when the momentum-relaxing collisions of electrons with impurities, phonons, and the boundary are significantly slower than the momentum-conserving electron-electron collisions.
Experimental evidence and observations of hydrodynamic flow in condensed matter systems has been steadily increasing in recent years across a handful of systems, such as two dimensional electron gases (2DEG) (Al,Ga)As~\cite{jong1995hydrodynamic}, graphene~\cite{Bandurin2016negative,Crossno2016observation,KrishnaKumar2017superballistic,Ku2020,Sulpizio2019}, and some bulk (semi)metals with nontrivial electronic structures~\cite{Gooth2018thermal,vool2020imaging,Nandi2018unconventional,moll2016evidence,narang2021topology}.

\begin{figure*}
    \centering
    \includegraphics[width=\linewidth]{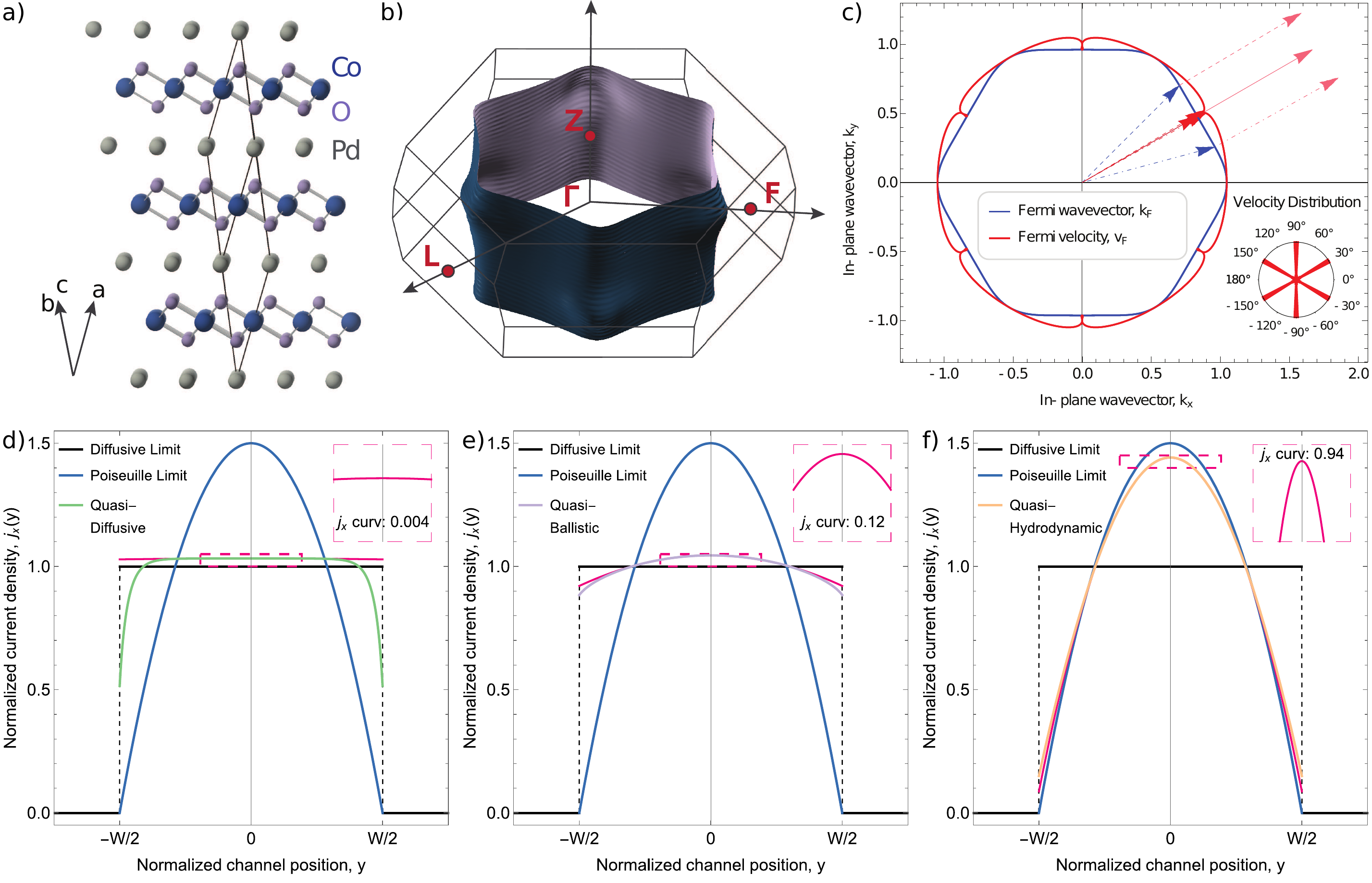}
    \caption{\textbf{(a)} Rhombohedral lattice structure of PdCoO$_2$ highlighting the vdW layers. 
    \textbf{(b)} Fermi surface of PdCoO$_2$ in the first Brillouin zone, with the high symmetry points denoted by red markers, showing an open faceted-cylindrical shape along the out-of-plane ($\Gamma-Z$) direction.
    \textbf{(c)} Polar plot of Fermi wavevector from ARPES measurements for PdCoO$_2$ (blue).
    Blue arrows plot the Fermi wavevector at $15^{\circ}$, $30^{\circ}$, and $45^{\circ}$.
    Pink arrows plot the vector perpendicular to the Fermi surface at those orientations, i.e. the Fermi velocity.
    Translated back to the zone origin, and plotted over all angles, red polar plot shows the angular dependence of the Fermi velocity.
    Note the nearly flat Fermi surface facets lead to a notable anisotropy in the velocity distribution (inset sector chart).
    Area-normalized current density profiles in a channel of width $W$ for \textbf{(d)} ``quasi-diffusive'', \textbf{(e)} ``quasi-ballistic'', and \textbf{(f)} ``quasi-hydrodynamic'' regimes.
    \textbf{(d-f)} also plot the fully-diffusive (black) and fully-parabolic (blue) limits for reference.
    Insets plot the best fit parabola at the middle of the channel, from which normalized current density curvatures are extracted.
    }
    \label{fig:fig1}
\end{figure*}

Among these systems, layered delafossites, such as PdCoO$_2$ and PtCoO$_2$ (\cref{fig:fig1}a),  are highly conductive metals with open hexagonal Fermi surfaces (\cref{fig:fig1}b), which can host long electron mean free paths even at the room temperature~\cite{Daou2017,Hicks2012quantum}, making them ideal candidates to observe finite-size effects in electron transport.
Despite their simple electronic structure, many interesting transport signatures have been observed in these systems, including large magnetoresistance and deviations from Kohler’s relation at finite temperatures~\cite{Nandi2018unconventional,takatsu2013extremely}, super-geometric electron focusing~\cite{bachmann2019super}, commensurability oscillations~\cite{putzke2020h}, upturn of in-plane resistivity~\cite{Hicks2012quantum}, directional ballistic effects~\cite{bachmann2021directional}, deviation between optical and transport scattering rate measurements at low temperatures~\cite{homes2019perfect}, and phonon drag effect in thermopower measurements~\cite{daou2015large}.
Curiously, the thermopower of PdCoO$_2$ was predicted to have the opposite signs along the in-plane and out-of-plane directions~\cite{Ong2010unusual,Daou2017}. 
Such axis-dependent $p-$ and $n-$type conduction has recently been proposed to originate from the open concave Fermi surface shape~\cite{He2019,wang2020anisotropic,Wang2020chemical}.
The emergence of such rich transport phenomena from a relatively simple electronic structure, poses many open questions.
For example, the extremely long carrier mean free path at low temperatures~\cite{Hicks2012quantum,sunko2020controlled}, combined with the recent observation of directional ballistics~\cite{bachmann2021directional}, has rekindled discussions on the hydrodynamic-character of electron transport in delafossite metals~\cite{moll2016evidence,cook2019polygonal,bachmann2021directional}.

In this \emph{Letter}, we address these unconventional electron transport phenomena, from a two dimensional Fermiology perspective, using PdCoO$_2$ as a model system. 
We establish the microscopic origins of electron scattering, from first principles, including momentum-relaxing events (electron-phonon and electron-impurity scattering) and momentum-conserving events (electron-electron scattering through the direct Coulomb interaction and exchnage of a virtual phonon).
We present these temperature-dependent electron mean free paths and discuss the dominant phonon dynamics arising from phonon-phonon and phonon-electron interactions.
We find that the phonon-mediated electron-electron interactions dominate the momentum-relaxing mechanisms at a wide temperature range, opening up the possibility for observing hydrodynamic electron transport regime.
Further, we explicitly examine the effects of Fermi surface anisotropy in the Boltzmann transport equation (BTE), which we use to extract the electron current flows in different transport \emph{limits}.
Using the microscopic length scales from the first principle calculations, we elucidate how the electron scattering physics manifest in unconventional transport phenomena in micro-scale samples.
Building on this, we plot the current profile curvature and in-plane conductivity for a range of momentum-relaxing and momentum-conserving interactions, allowing for classification of different transport regimes.
This work establishes a theoretical framework to understand electron transport in PdCoO$_2$ by providing first-principles insights into microscopic electron scattering mechanisms and bridging them with the Fermi surface geometry.
The approach demonstrated here is applicable to other condensed matter systems, including Dirac/Weyl semimetals, in predicting the emergence of hydrodynamics in these systems and distinguishing it from other transport regimes.

\textit{Anisotropic BTE and electron transport limits.---}
The evolution of a non-equilibrium charge carrier distribution function at steady-state and in the presence of an external electric field $\boldsymbol{E}$ is given by the semi-classical BTE:
\begin{align}
    \boldsymbol{v}_s \cdot \nabla_{\boldsymbol{r}}f(s,\boldsymbol{r}) + q \boldsymbol{E} \cdot \nabla_s f(s,\boldsymbol{r}) = \Gamma_s[f], \label{eq:BTE-1}
\end{align}
where $f(s,\boldsymbol{r})$ is the non-equilibrium charge carrier distribution around position $\boldsymbol{r}$ with state label $s$ (encapsulating the wavevector $\boldsymbol{k}$ and band index $n$), $\boldsymbol{v}_s$ is the carrier's group velocity, and $\Gamma_s[f]$ is the collision-integral functional.
We consider a two-dimensional infinite channel of width $W$, $\boldsymbol{r}=(x,y)$, and parametrize the single-band, $n=1$, in-plane wavevector at the Fermi surface using $\boldsymbol{k}=k_F\rho(\theta) \begin{Bmatrix}\cos(\theta) \\ \sin(\theta) \end{Bmatrix}$, where $k_F$ is the average Fermi wavevector and $\rho(\theta)$ is a polar equation defined between $\theta=0$ and $\theta=2\pi$, with an average value of 1.
Since the drift velocity along the channel's transverse direction must vanish at steady-state (see Supplemental Information Sec.~1), we linearize~\cref{eq:BTE-1} by introducing the parametrization
\begin{align}
    f(s,\boldsymbol{r}) = f(\theta,y) = q E_x v_x(\theta) l_{\mathrm{eff}}(\theta,y),
\end{align}
where $l_{\mathrm{eff}}$ is an effective distance a charge carrier at $y$ with momentum in the direction of $\theta$ has traveled since its last collision.
Under the relaxation time approximation, this leads to the integro-differential equation~\cite{jong1995hydrodynamic,vool2020imaging}:
\begin{align}
    v_y(\theta) \,\partial_y \, l_{\mathrm{eff}}(\theta,y) &+ \frac{l_{\mathrm{eff}}(\theta,y)}{l} \notag \\ 
    &= 1 + \frac{1}{l_{\mathrm{mc}}}\int_0^{2\pi} \frac{d \phi}{\pi} v_x^2(\phi) l_{\mathrm{eff}}(\phi,y),
\label{eq:BTE-2}
\end{align}
where $l_{\mathrm{mc}}$ is a momentum-conserving mean-free path, and we've used Mathiessen's rule $l^{-1}=l_{\mathrm{mr}}^{-1}+l_{\mathrm{mc}}^{-1}$ to combine $l_{\mathrm{mc}}$ with the momentum-relaxing mean free path $l_{\mathrm{mr}}$.
The Fermi-surface anisotropy is introduced through the Fermi velocity term 
\begin{align}
    \boldsymbol{v}=\begin{Bmatrix}\cos(\theta)\rho(\theta) + \sin(\theta)\rho'(\theta) \\ \sin(\theta)\rho(\theta) - \cos(\theta)\rho'(\theta) \end{Bmatrix} \label{eq:vel}.
\end{align}

The in-plane Fermi surface of PdCoO$_2$, based on Angle-resolved photoemission spectroscopy (ARPES) measurements~\cite{bachmann2021directional}, is shown in~\cref{fig:fig1}c.
While the hexagonal Fermi surface does not deviate strongly from an isotropic (circular) Fermi surface, its faceted nature leads to a highly anisotropic velocity distribution (\cref{fig:fig1}c).

Before examining the effects of this velocity anisotropy on~\cref{eq:BTE-2}, it is instructive to clarify \emph{limits} arising in electron transport regimes, namely the common diffusive or ``Ohmic'' regime and the more exotic hydrodynamic or ''Poiseuille'' limit.
These limits arise in wide-enough channels when one of the momentum-relaxing or momentum-conserving collisions dominate over the other and result in completely uniform or completely parabolic current profiles for ``Ohmic'' (\cref{fig:fig1}d-f black) and ``Poiseuille'' (\cref{fig:fig1}d-f blue) flows respectively.
In finite-size conductors, these scattering length scales additionally compete with the channel's boundary, giving rise to ballistic transport, where the scattering with the sample wall is the dominating mechanism.

Given that in high quality crystals momentum-relaxing mean free paths are on the order of microns, electron transport regimes in micro- and nanoscale devices exist somewhere on a spectrum between the ``Ohmic'', ``Poiseuille'', and ballistic limits. 
\Cref{fig:fig1}d-f plots such current profiles, which we term ``quasi-diffusive'', ``quasi-ballistic'', and ``quasi-hydrodynamic'' to further highlight the continuous transition between these limiting regimes.
It is worth emphasizing that, given that the quasi-hydrodynamic regime arises mostly from momentum-conserving scattering events, which cannot be readily captured using traditional transport measurements such as resistivity, experimentally detecting the manifestation of such current profiles is not trivial.

Recently, spatially resolved measurements have captured non-uniform current density distributions in graphene~\cite{Sulpizio2019,Ku2020,jenkins_imaging_2020} and WTe$_2$~\cite{vool2020imaging} conducting channels.
As such, we use the calculated current density distribution, and specifically its curvature at the middle of the channel, as an important metric to discuss transport phenomena in finite-size PdCoO$_2$. 
Finally, we note that hydrodynamic effects are significant when the hierarchy of the electron length scales due to scattering events, including momentum-relaxing scattering with phonons and impurities ($l_{\mathrm{mr}}$), momentum-conserving scattering with other electrons ($l_{\mathrm{mc}}$), as well as scattering with the boundary of the channel ($W$) satisfies the inequalities $l_{\mathrm{mc}}\ll W\ll l_{\mathrm{mr}}$~\cite{cook2019polygonal,LucasKCFong2018hydrodynamic}.
This naturally leads to the observation that ballistic and hydrodynamic effects can coexist and contribute to the overall transport measurements.

\begin{figure*}
    \centering
    \includegraphics[width=\linewidth]{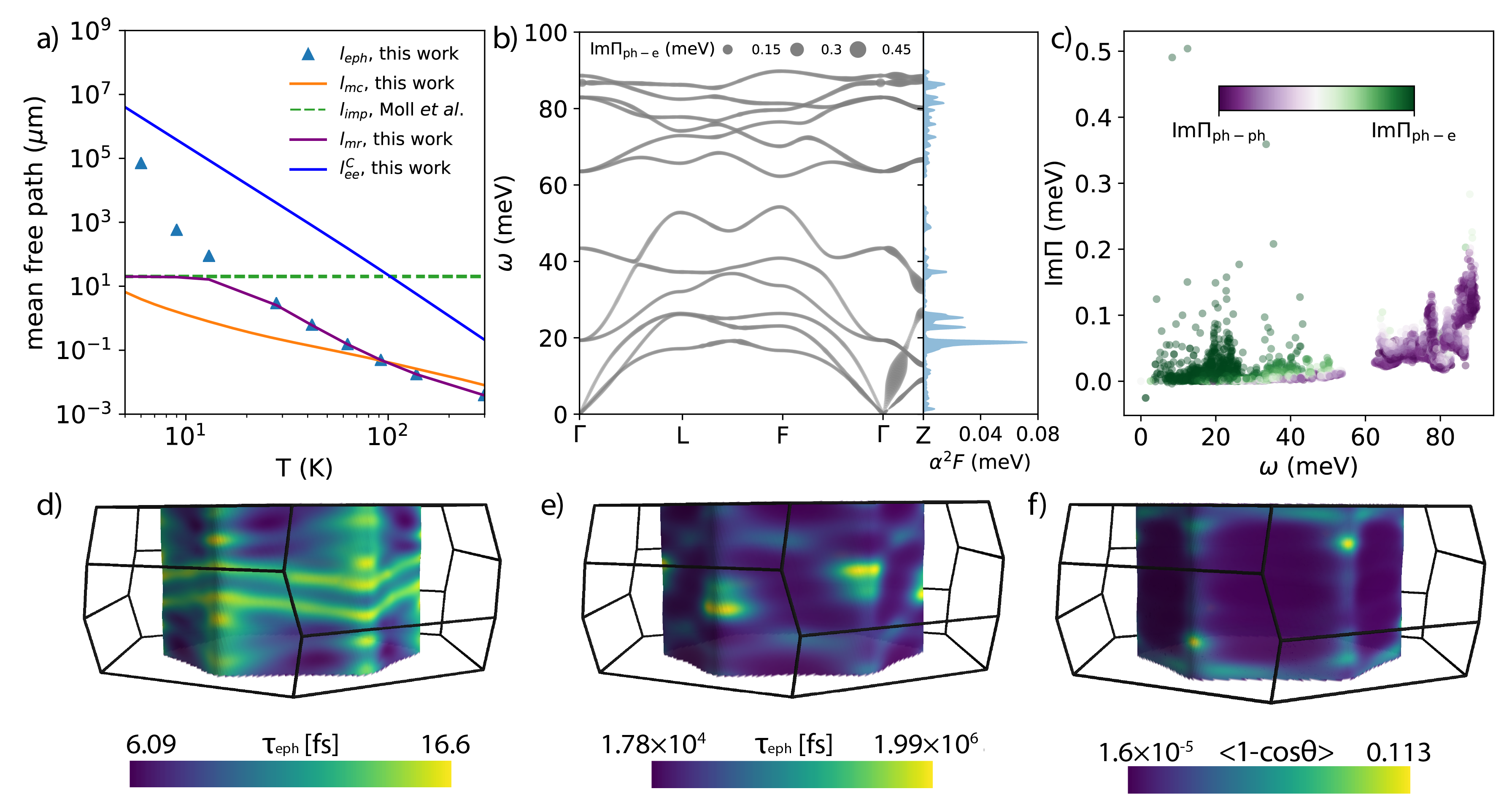}
    \caption{\textbf{(a)} Temperature dependent electron mean free pahts for PdCoO$_2$, obtained from \emph{ab initio} calculations (see Methods). We consider electron-phonon ($l_{\mathrm{eph}}$), electron-impurity ($l_{\mathrm{imp}}$), electron-electron scattering mediated by a screened Coulomb interaction ($l_{\mathrm{ee}}^C$), and by a virtual phonon ($l_{\mathrm{mc}}$).
    The momentum relaxing mean free path, $l_{\mathrm{mr}}$, is computed using Matthiessen's rule.
    \textbf{(b)} Phonon dispersion of PdCoO$_2$ along the high symmetry path in the Brillouin zone (\cref{fig:fig1}b), with the marker size indicating the imaginary part of phonon-electron self energy ($\rm{Im}\Pi{\rm ph-e}$) calculated at 25~K.
    The corresponding Eliashberg coupling function for different phonon frequencies, $\alpha^2F(\omega)$ (see Methods), is plotted in the right panel. 
    \textbf{(c)} Comparison of phonon self energies at 25~K resulting from phonon-phonon ($\rm{Im}\Pi_{\rm ph-ph}$) and phonon-electron coupling ($\rm{Im}\Pi_{\rm ph-e}$), highlighting that the phonon-electron interaction dominates the low energy modes at low temperatures. 
    Anisotropic scattering distribution on the Fermi surface, illustrating $\tau_{\rm eph}(n{\bf k})$ at \textbf{(d)} 300~K, \textbf{(e)} 4~K, and \textbf{(f)} the scattering efficiency $<1-{\rm cos}\theta >$ at 4~K.}
    \label{fig:fig2}
\end{figure*}

\textit{Microscopic scattering mechanisms in {P\lowercase{d}C\lowercase{o}O$_2$}.---}
As previously discussed, it is challenging to distinguish transport signatures between the quasi-hydrodynamic and quasi-ballistic regimes. 
Consequently, it is important to address the microscopic mechanisms giving rise to each regime using first-principle calculations of electron lifetimes due to scattering events specific to the materials in question.
For instance, in WTe$_2$ phonon-mediated momentum-conserving electron interactions dominate momentum-relaxing interactions of electrons with phonons and impurities at intermediate temperatures and thus lead to hydrodynamic transport flow~\cite{vool2020imaging}. 
This scattering mechanism is often overlooked in studies of low-carrier density materials, where electrons scatter predominantly  through the more-direct Coulomb interaction.
By contrast, bulk PdCoO$_2$ has a highly dispersive band with considerable carrier concentration at the Fermi level, where the Coulomb interaction is expected to be screened contributing weakly to electron-electron scattering.

\Cref{fig:fig2}a plots the temperature dependent electron mean free paths computed for various interactions using first principles (see Supplemental Information Sec.~2).
Here, momentum relaxing ($l_{\mathrm{mr}}$) events include the electron-phonon scattering ($l_{eph}$), which decreases rapidly with temperature and the electron-impurity scattering ($l_{imp}$), which is largely temperature independent and dominated by the impurity concentration in the sample.
Momentum conserving events ($l_{\mathrm{mc}}$) include the electron-electron scattering mediated by the direct Coulomb interaction ($l_{ee}^C$), as well as by a virtual phonon.
Below $\sim10$~K, the long electron-phonon mean free path in PdCoO$_2$ leaves the resistivity fully determined by the impurity length.
Particularly, the as-grown samples are shown to be resistant to defect formation~\cite{sunko2020controlled}, and the resulting long $l_{\mathrm{mr}}$ provides a temperature-window for $l_{\mathrm{mc}}$ to dominate the dynamics.
Similar to observations in WP$_2$~\cite{coulter2018microscopic} and WTe$_2$~\cite{vool2020imaging}, the Coulomb-mediated electron-electron mean free paths are orders of magnitude longer than those associated with the other mechanisms, due to their high carrier density.
Importantly, the phonon-mediated electron mean free paths in PdCoO$_2$ are significantly lower than $l_{\mathrm{mr}}$ below $\sim 100$~K, indicating the possibility for quasi-hydrodynamic flow when the channel width satisfies $l_{\mathrm{mc}} \ll W \ll l_{\mathrm{mr}}$.

To understand the role of phonons in the electronic transport through this phonon mediated interaction, we investigate the phonon dynamics by comparing the phonon self-energies due to phonon-electron and phonon-phonon interactions. 
\Cref{fig:fig2}b plots the imaginary part of the phonon self-energy (Im$\Pi_{\rm ph-e}$) due to phonon-electron interactions for different phonon modes (left panel).
The phonon dispersion is calculated along the high-symmetry lines in the Brillouin zone illustrated in Fig.~\ref{fig:fig1}b, where $\Gamma$-F and $\Gamma$-L represent the in-plane direction while $\Gamma$-Z represents the cross-plane direction along the stacking of the [CoO$_2$]$^{-1}$ layers.
The cross-plane phonon modes show much larger self-energies than the in-plane ones, for both acoustic and optical branches, indicating these modes are more strongly coupled with electrons.
The Eliashberg spectral function, plotted at different phonon frequencies (Fig.~\ref{fig:fig2}b: right panel), indicates that the lower energy optical phonon modes contribute the most.

Further, since the phonon-mediated electron electron interaction proceeds via the exchange of a `virtual` phonon, i.e. the phonon emitted (absorbed) by a pair of electrons is assumed to by `instantaneously' absorbed (emitted) by a different pair of electrons, we explicitly examine the phonon self-energies due to the competing phonon-phonon interactions (Im$\Pi_{\rm ph-ph}$) to justify this assumption.
Fig.~\ref{fig:fig2}c compares the two phonon scattering mechanisms at 25~K, well within the temperature window identified in Fig.~\ref{fig:fig2}a for manifesting quasi-hydrodynamic flow.
While the high energy optical modes feature stronger phonon-phonon scattering, the lower energy ones show much stronger phonon-electron interactions.
As such, at low temperatures where only lower-energy phonon modes are occupied, the phonon linewidths follow $\mathrm{Im\Pi_{ph-ph}}\ll \mathrm{Im\Pi_{ ph-e}}$, indicating that phonons rapidly exchange their momentum with electrons instead of relaxing their momentum to the lattice, i.e. thermalizing.
As the lattice temperature increases, phonon-phonon interactions increase and finally overtake the phonon-electron coupling (fig.~S1).
Recently, the unconventional temperature-dependent phonon linewidth decay observeved in a Raman study in WP$_2$~\cite{Osterhoudt2011evidence}, was also attributed to the strong phonon-electron coupling leading to the similar hierarchy of the quasiparticle lifetimes.

\begin{figure*}
    \centering
    \includegraphics[width=\linewidth]{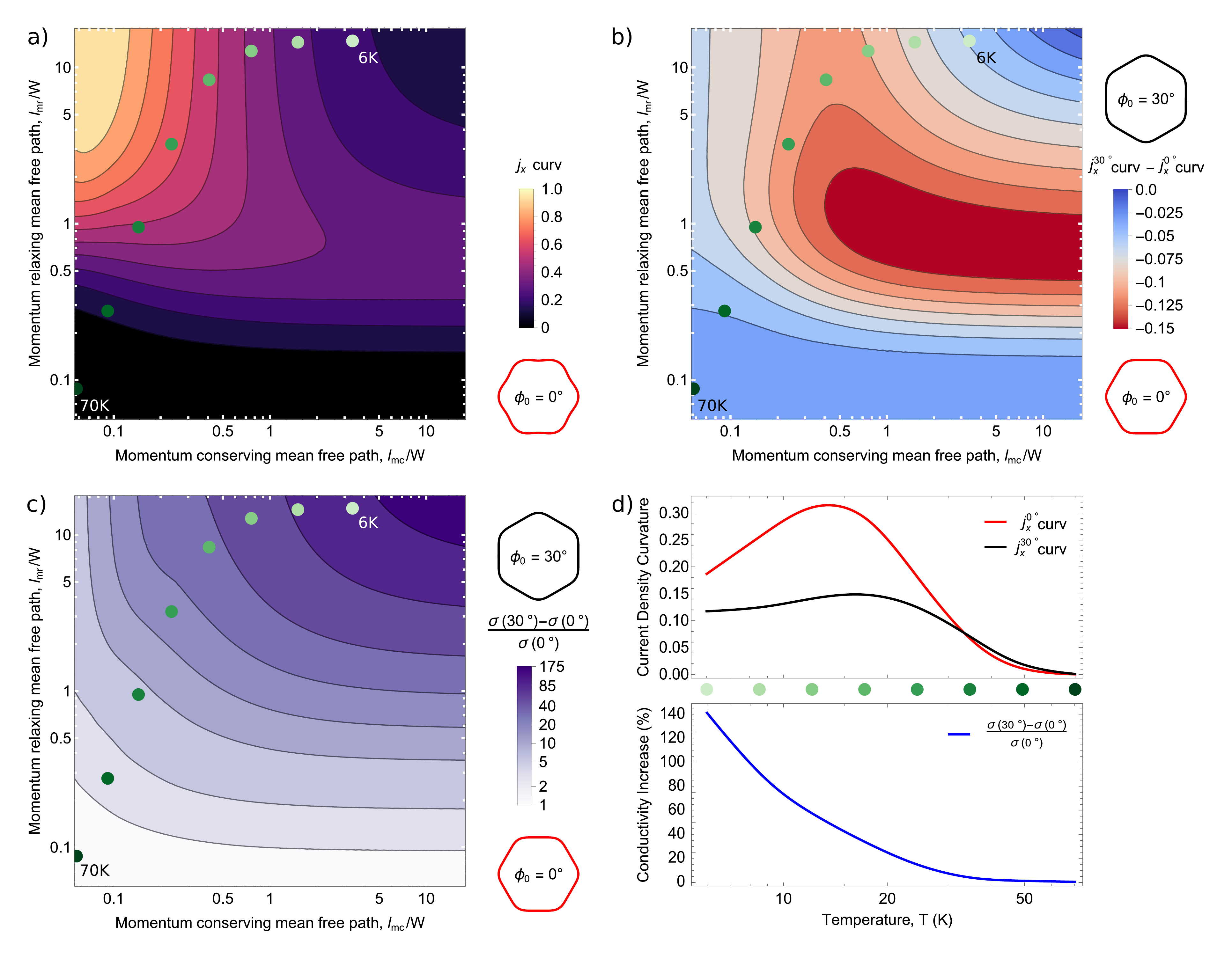}
    \caption{\textbf{(a)} Current density curvature contour plot over a large range of momentum-conserving ($l_{\mathrm{mc}}/W$, x-axis) and momentum-relaxing ($l_{\mathrm{mr}}/W$, y-axis) interactions, for a system with Fermi surface shown in red in the inset.
    Overlaid points plot the \textit{ab initio} temperature-dependent mean free paths for a 1.25 $\mu m$ wide channel of PdCoO$_2$ with an impurity level of 18.5 $\mu m$; temperature range is 6K (lightest green) to 70K (darkest green).
    \textbf{(b)} Current density curvature difference between the system shown in (a) and the same with a rotated Fermi surface at $30^{\circ}$ with respect to the channel (black inset).
    The decrease in curvature is indicative of the shift from ``quasi-hydrodynamic'' to ``quasi-ballistic'' flow.
    \textbf{(c)} Percent difference in in-plane conductivity between the two Fermi surface orientations showin in (b).
    \textbf{(d)} Current density curvature (top) and percent difference in conductivity (bottom) along the \textit{ab initio} temperature trajectory shown by the overlaid green points in (a-c).
    While the directional dependence naturally manifests itself most prominently in the ``quasi-ballistic'' regime, it is still evident in the ``quasi-hydronamic'' window between 10-25K.}
    \label{fig:fig3}
\end{figure*}  

In the absence of electron-impurity scattering, PdCoO$_2$ exhibits extremely long electron-phonon scattering mean free paths (\cref{fig:fig2}a blue markers).
To better understand the origin of these long-lived electrons, we computed the state-resolved electron-phonon lifetimes ($\tau_{\rm eph}$) for the single energy band.
For transport properties, only the states (with energy $\varepsilon_{\vec{k}}$) near the Fermi level matter, typically weighted by $-f'(\varepsilon_{\vec{k}})$ in the Boltzmann relaxation time approximation (RTA).
The distribution of the strongly anisotropic $\tau_{\rm eph}$ on the hexagonal Fermi surface is displayed in~\cref{fig:fig2}d-e for high (300~K) and low (4~K) temperatures, with yellow color corresponding to ``long-lived'' carriers.
While at higher temperatures, the long $\tau_{\rm eph}$ is more-evenly distributed between the faces and the edges of the hexagonal Fermi surface, the low temperature calculations show a much sharper contrast, whereby ``long-lived'' carriers are only present in regions of the Fermi surface with large Gaussian curvature.
Further, while the difference between ``long-'' and ``short-lived'' carriers is within a factor of three at 300~K, it spans more than two orders of magnitude at 4~K.
Further, we augment the scattering rate with an efficiency factor given by the relative change in direction of the initial and final state velocities $\left(1-\frac{v_{n {\bf k}}\cdot v_{m{\bf k'}}}{|v_{n {\bf k}}|| v_{m{\bf k'}}}| = 1-\cos(\theta)\right)$~\cite{ziman2001electrons}, where $\theta$ is the scattering angle between the initial and final electronic states.
Phonon modes strongly coupled with electrons appear to have a small momentum transfer {\bf q} at very low temperatures, resulting in the vanishing scattering efficiency, shown by $<1-{\rm cos}\theta>\approx 0$ in~\cref{fig:fig2}f.
Further, ``long-lived'' carriers at low temperatures correlate with regions of larger negative Gaussian curvature(\cref{fig:fig1}).
This can give rise to rich transport phenomena in PdCoO$_2$ micron-scale devices as we discuss in greater detail below, as the momentum relaxing electron mean free path can reach $\sim20\mu{\rm m}$ at low temperatures even in the presence of impurities~\cite{moll2016evidence,bachmann2021directional}.

\textit{Electron transport regimes.---}
Equipped with the temperature-dependent momentum-relaxing and momentum-conserving mean free paths for PdCoO$_2$ using first-principles, we solve~\cref{eq:BTE-2} for a range of $\left(l_{\mathrm{mr}}/W,\; l_{\mathrm{mc}}/W\right)$ pairs to examine the dominating transport regimes at different temperatures.
\Cref{fig:fig3}a shows a contour plot of the curvature of the current profile at the middle of the channel for a hexagonal Fermi surface aligned at $0^{\circ}$ with the channel, with the ``quasi-diffusive'', ``quasi-ballistic'', and ``quasi-hydrodynamic'' regimes corresponding to the bottom right ($l_{\mathrm{mc}} \gg W, l_{\mathrm{mr}} \ll W$), top right ($l_{\mathrm{mc}}, l_{\mathrm{mr}} \gg W$), and top left ($l_{\mathrm{mr}} \gg W, l_{\mathrm{mc}} \ll W$) respectively.
The overlaid green points mark the \textit{ab initio} temperature trajectory~\cref{fig:fig2}a, with lighter color indicating lower temperature. 
Due to the large ratio of $l_{\mathrm{mr}}$ to $l_{\mathrm{mc}}$, the phonon-mediated electron interactions lead to a strongly non-uniform profile between 10-25K.
To address the anisotropy of the hexagonal Fermi surface shape, we compare the same calculation for a rotated Fermi surface aligned at $30^{\circ}$ with the channel, shown in \cref{fig:fig3}b.
We observe a considerable decrease in the observed current profile curvature, indicating a transition away from quasi-hydrodynamic to quasi-ballistic behavior.
Intuitively, this directional dependence peaks in the ``quasi-ballistic'' regime as the boundary scattering dominates, and vanishes in the ``quasi-diffusive'' limit as the multiple scattering events serve to effectively randomize the carriers' velocity directions.

Moreover, such directional dependence of the current flow is strongly manifested in the device's in-plane conductivity (\cref{fig:fig3}c), which is directly observable from transport measurements.
This anisotropic effect is strongest in the absence of momentum-conserving interactions (i.e. at the far-right of~\cref{fig:fig3}), indicating the emergence of `easy' and `hard' transport directions~\cite{bachmann2021directional}.
As momentum-conserving interactions lead to collective flow which likewise acts to homogenize the carriers' velocity directions, the effect gets weaker in the quasi-hydrodynamic regime identified between 10-25~K, albeit still expected to be measurable.
\Cref{fig:fig3}d plots the \% increase for both current density curvatures (top) and the in-plane conductivity (bottom) as a function of temperature, the latter of which has been recently observed in Ref.~\cite{bachmann2021directional,bachmann2019super}.
These transport regime plots effectively bridge together the microscopic scattering events and the exotic transport signatures in PdCoO$_2$ and highlight how the Fermi surface topology could give rise to behaviors deviating from the simple kinetic picture.

\textit{Conclusions and outlook.---}
We present an investigation of unconventional electron transport in the delafossite metal PdCoO$_2$, using a combination of first-principles methods and numerical modeling of the anisotropic BTE.
We examine 
the temperature-dependent electron mean free paths, with an emphasis on establishing the hierarchy of the momentum-relaxing ($l_{\mathrm{mr}}$) and momentum-conserving ($l_{\mathrm{mc}}$) interactions.
We show that while the direct Coulomb interaction is largely screened due to the metal's high carrier density, the phonon-mediated electron-electron interaction dominates momentum-relaxing electron interactions in the temperature range of $4-100$~K.
This mechanism is supported by the much smaller phonon-phonon self-energies compared to the phonon-electron interaction at low temperatures.
We develop the formalism for an anisotropic BTE which treats the ARPES-measured in-plane Fermi surface shape, and obtain the numerical solutions for a wide range of $l_{\mathrm{mr}}/W$, $l_{\mathrm{mc}}/W$ pairs.
We discuss the resulting transport regime \emph{limits} and demonstrate the directional dependence of these transport measurements with regards to the orientation of the Fermi surface to the conducting channel.

Our work offers insight into previous observations of in PdCoO$_2$, and presents a comprehensive theoretical framework to distinguish the quasi-ballistic and quasi-hydrodynamic transport regimes. 
The strong phonon-mediated electron interactions suggest the possibility of hydrodynamics in PdCoO$_2$, which we anticipate will spark experimental investigation via spatially resolved techniques.
Finally, our findings serve as a foundation for understanding electron transport from microscopic scattering mechanisms and Fermi surface geometry, and motivate future efforts in computing the electronic viscosity in such material systems directly from first principles.
\newline
\newline
\indent
The authors acknowledge fruitful discussions with Adam Jermyn, Uri Vool, Claudia Felser, Doug Bonn, Johannes Gooth, and Amir Yacoby.
This work was supported by the Quantum Science Center (QSC), a National Quantum Information Science Research Center of the U.S. Department of Energy (DOE). Y.W. was partially supported during the project by the STC Center for Integrated Quantum Materials, NSF Grant No. DMR-1231319 for development of computational methods for topological materials. P.N. is a Moore Inventor Fellow and gratefully acknowledges support through Grant No. GBMF8048 from the Gordon and Betty Moore Foundation. P.J.W.M. acknowledges support by the Swiss National Science Foundation (176789). This research used resources of the Oak Ridge Leadership Computing Facility, which is a DOE Office of Science User Facility supported under Contract DE-AC05-00OR22725 as well as the resources of the National Energy Research Scientific Computing Center, a DOE Office of Science User Facility supported by the Office of Science of the U.S. Department of Energy under Contract No. DE-AC02-05CH11231.

\bibliographystyle{apsrev4-2}
\bibliography{PdCoO2-citations}

\end{document}